\documentstyle[aps,preprint]{revtex}

\title{The effective action for a relativistic Jaynes-Cummings model}

\author{Mark Burgess\\Faculty of Engineering, Oslo College, 0254 Oslo,
Norway\\and\\Institute of Physics, University of Oslo, P.O.Box 1048 Blindern,
0316 Oslo, Norway\\~\\ Margaret Carrington and Gabor Kunstatter\\Department of
Physics and Winnipeg Institute for Theoretical Physics\\
 University of Winnipeg, Winnipeg Manitoba, Canada R3B 2E9}
\date{\today}

\date{\today}

\def\boxx{\vcenter{\vbox{\hrule height.4pt
          \hbox{\vrule width.4pt height8pt
          \kern8pt\vrule width.4pt}\hrule height.4pt}}}

\begin{document}
\bibliographystyle{unsrt}
\newcommand{\be}{\begin{equation}}
\newcommand{\ee}{\end{equation}}
\newcommand{\bea}{\begin{eqnarray}}
\newcommand{\eea}{\end{eqnarray}}

\maketitle

\begin{abstract}
We describe an effective field theory for atomic lasers which reduces
to the Jaynes-Cummings model in the non-relativistic, single mode
limit.  Our action describes a multi-mode system, with general
polarizations and Lorentz invariance and can therefore
be used in all contexts from the astrophysical to the laboratory. We
show how to compute the effective action for this model and perform
the calculation explicitly at the one loop level.  Our model provides a way of
analyzing a many-particle, two-state model with arbitrary boundary
conditions.
\end{abstract}

\section{Introduction}
The canonical model for laser physics is the Jaynes-Cummings model.
It describes a single mode oscillator representing a coherent
electromagnetic field, coupled to a two level reservoir of
atoms\cite{jaynes1}. The Jaynes-Cummings model is defined in momentum
space in terms of the photon creation and annihilation operators $a$ and
$a^\dagger$ for a single momentum mode $K = \Omega/c$ and a single, unspecified
polarization of the electromagnetic field. The quantum mechanical
Hamiltonian is given by
\begin{equation}
H = a^\dagger a \hbar\Omega + \frac{1}{2}\hbar \omega_{12} \sigma_z
 + \hbar g (\sigma_+ a + a^\dagger \sigma_-),
\label{eq:1}
\end{equation}
where $\hbar \omega_{12}$ is the energy difference between the atomic
states. The creation and annihilation operators satisfy
$[a,a^\dagger]=1$ and the sigma matrices satisfy
$[\sigma_+,\sigma_-]=\sigma_z$.  This model is the natural candidate
for studying the fundamentals of the interaction between matter and
radiation in a laser in a wide variety of situations, but it has
several shortcomings and it is important to understand how these may
be resolved in a reasonable fashion.

In this paper we present a new model which retains the essential
simplicity of the Jaynes-Cummings model, but which repairs some of its
limitations.  One of our principal aims is to write down a real-space
Lagrangian formulation for a two-state system in which spectral content
and polarizations are fully general: this should not only allow us to
use the full machinery of modern field theory with all its attendant
methodology, but also admit the solution of problems with general
boundary conditions, in contact with many particle reservoirs. The
theory makes gauge symmetries and the space-time structure clearer and
leaves us free to use well-established path integral or Green function
methods for computing the effective action.  Finally, but not
least importantly, it also bridges a cultural
gap between the worlds of field theory and laser physics.

The Jaynes-Cummings model is an idealized description of laser
phenomena. As a single-mode theory it can not address boundary
conditions\cite{parker1,cook1} or time-dependent
interactions\cite{burgess13} since, by the uncertainty principle, a
single mode must be completely delocalized in space and time. These
are features characteristic of the micro-maser and of non-linear
media. Almost all of the work on lasers is phenomenological and
couched in momentum space. Real space methods were pioneered by Graham
and Haken\cite{graham1,graham2,graham3}, but the closest attempts at
constructing a microscopic description of the laser come from
Korenman's\cite{korenman1} use of Schwinger's action
formulation\cite{schwinger2}. A recent letter makes some progress with
this approach for semiconductor lasers\cite{henneberger1}. These
papers also deal with effective theories however.  Our paper is no
different in this respect: we use an effective interaction and
effective field variables.  Indeed it would be inconceivable to
attempt to write down a theory in which every optically-active
electron and background charge were dealt with explicitly.  Rather we
pose the question: what are the relevant degrees of freedom for the
laser at the energy scales of interest? These are clearly the
averaged atomic properties and the magnitude of electromagnetic field.

A disadvantage with Korenman's analysis is his use of non-relativistic
field theory. Korenman begins with the Schr\"odinger equation coupled
to reservoirs and seeks self-consistent solutions for decays
rates and line widths. But radiative corrections to the
non-relativistic theory are beset with problems: acausal loop
diagrams, such as those used in constructing the effective action,
vanish owing to the absence of anti-particles (negative energy states)
in the non-relativistic theory. This makes the non-relativistic theory
alien to field theorists who are used to the language of Feynman
diagrams and Green functions and, in any case, one would expect a
physical system described by the Schr\"odinger equation to arise
naturally from a more general relativistic theory in the low energy
limit.  There is then the issue of non-renormalizability:
Schr\"odinger scalar field theory is more divergent than relativistic
scalar field theory, owing to the dimension of the field variables,
and is specifically non-renormalizable in $3+1$ dimensions. It
therefore makes more sense to begin with a relativistic theory,
which is renormalizable, and consider the non-relativistic theory as
an approximation to this full theory.  In addition, we expect that a
relativistic theory is necessary to study astrophysical situations, where
the motion of atoms could be relativistic at high temperatures, even
when the emitted radiation is of low energy. We begin therefore by
introducing the action for a relativistic two-state model.

\section{The action and its interpretation}

Consider a system of neutral atoms, containing optically-active
electrons, which endow the atoms with a dipole moment.  The electrons
will not be explicit degrees of freedom in our model, rather their
presence will be taken into account by the availability of transitions
between the two atomic states.  A neutral atoms is therefore
represented as a two component real-scalar field; the two components
represent the lower (unexcited) and upper (excited) levels of the
atom. Each level has a different effective `mass', in relativistic
terminology $m_{a} = m +E_{a}/c^2$, where $m$ is the atomic mass and
the potential-energy of the level is $E_a$. In SI units, the action
has the following form:
\begin{equation}
S = \int dV_x
\left\{\frac{1}{2}\hbar^2c^2(\partial^\mu\phi_{a})(\partial_\mu\phi_{a})
+\frac{1}{2}m_{a}^2 c^4 \phi_{a}\phi_{a} +
\frac{1}{4\mu_0}F^{\mu\nu}F_{\mu\nu}
+P^{\mu\nu}(\phi)F_{\mu\nu}  \right\}
\label{eq:2}
\end{equation}
where ${a}=1,2$ and $P^{\mu\nu}(\phi)$ is a polarization tensor which
is to be specified below.  Our conventions are such that the Minkowski
metric tensor $g_{\mu\nu}$ has the signature $-+++$ and we use symbols
$d\sigma_x$ to represent an $n$-dimensional infinitesimal spatial
volume element on a spacelike hypersurface and $dV_x$ to represent the
$n+1$-dimensional infinitesimal spacetime volume which is canonically
written $d\sigma_x dx^0 \sqrt{-{\rm det} g}$. In other words,
$\sigma_x$ is a spatial volume and $V_x$ is a spacetime volume.

The essential physics of this model is determined by the form of the
dipole interaction tensor $P^{\mu\nu}(\phi)$. Given that the dynamical
degrees of freedom are represented by real scalar fields, we have only
two choices for this quantity, as we discuss below. The form for
such a dipole term is unfamiliar in a relativistic theory, so we allow
ourselves to be guided by the non-relativistic limit and require that
this limit be consistent with known results, namely the non-relativistic
analysis of Korenman\cite{korenman1} and in turn the Jaynes-Cummings
model\cite{jaynes1}. In particular, in the non-relativistic limit,
one should obtain an expression for $P_{\mu\nu}$ of the form used by
Korenman:
\begin{equation}
P_{\mu\nu} \rightarrow \gamma_{\mu\nu}^{{a}{b}}\psi^*_{a}\psi_{b},
\label{eq:12}
\end{equation}
for some constant, off-diagonal matrix $\gamma_{\mu\nu}^{{a}{b}}$.
It is evident that this is a dipole induced transition from the form of
the operators. $\psi^*$ is a creation operator for the field and
$\psi$ is a destruction operator, thus the off-diagonal operator
creates an upper state and destroys a lower state, or vice-versa.
Moreover, the components $\gamma_{0i}$ of this matrix will be
proportional to the electric dipole moment of the atom. One
relativistic generalization which reduces to eqn. (\ref{eq:12}), is
\begin{equation}
\tilde P_{\mu\nu} = i \hbar\gamma_{\mu\nu}\tilde
\epsilon^{{a}{b}}\phi_{a}\partial_0\phi_{b},
\label{eq:13}
\end{equation}
where $\gamma_{\mu\nu}$ is a constant, anti-symmetric
tensor and $\tilde\epsilon_{{a}{b}}$ is the
two-dimensional antisymmetric Levi-Civita symbol\cite{jcfoot1}.
This form is intuitively appealing
because it seems to be related to the relativistic inner product:
\begin{equation}
(\phi_{a},\phi_{b}) = i \hbar c^2 \int d\sigma_x \frac{1}{2}
(\phi_{a}^*\partial_0\phi_{b}-(\partial_0\phi_{a}^*)\phi_{b}).
\label{eq:10}
\end{equation}
Unfortunately, this form for $\tilde P_{\mu\nu}$ raises some questions
concerning renormalizability
(see section V). It is non-renormalizable in $3+1$ and $2+1$
dimensions: in particular we expect new, higher derivative interactions to
be introduced at each order in perturbation theory. Although low energy
predictions are still possible in such theories, we avoid this problem
by introducing another interaction
\begin{equation}
 P_{\mu\nu} = {1\over 2}\gamma_{\mu\nu}\overline\epsilon^{{a}{b}}
\phi_{a}\phi_{b},
\label{eq:14}
\end{equation}
where $\gamma_{\mu\nu}$ is antisymmetric in $\mu$ and $\nu$ and
$$
\overline\epsilon^{{a}{b}}=
 \left(\begin{array} {cc}
  0 & 1 \\ 1 & 0
   \end{array}\right)
$$
 is now {\em symmetric} in ${a}$ and
${b}$.  In the non-relativistic limit,
this interaction differs from that in eqn. (\ref{eq:13}) only by a
factor of $i\hbar/mc^2$ (introduced by $\partial_t$). It has the
advantage of being marginally renormalizable in $3+1$ dimensions and
super-renormalizable in $2+1$ dimensions. This will be discussed in
more detail in Section V, in which the one loop effective action is
computed explicitly.

We now sketch a simple derivation of the non-relativistic limit for
our action, in order to give a physical interpretation to the constant
matrix $\gamma_{\mu\nu}$. The next section contains a further
justification of this method based on the field equations. The
limiting procedure is unambiguous up to redefinitions of the origin
for the arbitrary energy scale. The simplest
procedure is to first observe that the real scalar field $\phi(x)$ may
be decomposed into
\begin{equation}
\phi(x) = \phi^{(+)}(x) + \phi^{(-)}(x),
\label{eq:5}
\end{equation}
where $\phi^{(+)}(x)$ is the positive frequency part of the
 field and $\phi^{(-)}(x)$ is the negative frequency part
 of the field and $\phi^{(+)}(x)=(\phi^{(-)}(x))^*$. We now rescale
 the fields by the atomic mass:
\begin{equation}
\phi^{(+)}(x) = \frac{\psi(x)}{\sqrt{2mc^3}}~~ ,~~~~~ \phi^{(-)}(x)
=\frac{\psi^*(x)}{\sqrt{2mc^3}}
\label{eq:7}
\end{equation}
In addition, we note that the relativistic energy operator
$i\hbar\partial_t$ is related to the non-relativistic energy operator
$i\hbar\tilde\partial_t$ by a shift with respect to the rest energy of
particles:
\begin{equation}
i\hbar\partial_t = mc^2 + i\hbar\tilde\partial_t.
\label{eq:i1}
\end{equation}
This is because the non-relativistic Hamiltonian does not include the
rest energy of particles, its zero point begins just above the rest energy.

 Integrating the kinetic term by parts so
that $(\partial_\mu\phi)^2\rightarrow
\phi(-\boxx)\phi$ and substituting eqn. (\ref{eq:7}) into eqns. (\ref{eq:2})
and (\ref{eq:14})
gives,
\begin{eqnarray}
S &=& \int d\sigma_x dt \frac{1}{2}(\psi+\psi^*)_{a}\left\{
\frac{\hbar^2\tilde\partial_t^2}{mc^2} -i\hbar\tilde\partial_t +
\frac{E_{a}^2}{2mc^2}
+ E_{a} -\frac{\hbar^2 }{2m} \nabla^2
\right\}(\psi+\psi^*)_{a} \nonumber\\
&+& \int d\sigma_x dt \frac{\gamma^{\mu\nu}\overline \epsilon_{{a}{b}}}{4mc^2}
F_{\mu\nu}(\psi+\psi^*)_{a}(\psi+\psi^*)_{b}\label{eq:i2}.
\end{eqnarray}
Here we have dropped the Maxwell part of the action to avoid clutter,
since it has no non-relativistic limit.
If we use the fact that $\psi_{a}(x)$ is composed of only positive
plane-wave frequencies, it follows that terms involving $\psi^2$ or
$(\psi^*)^2$ vanish since they involve delta functions imposing a
non-satisfiable condition on the energy
$\delta(mc^2+\hbar\tilde\omega)$, where both $m$ and $\tilde\omega$
are greater than zero.  This assumption ceases to be true only if
there is an explicit time-dependence in the action, indicating a
non-equilibrium scenario, or if the mass of the atoms goes to zero (in
which case the NR limit is unphysical).  In the next section we
perform a transformation of the field equations which decouples the
positive and negative frequency modes, justifying this procedure in a
more conventional way.  We are therefore left with
\begin{equation}
S_{NR} =  \lim_{c\rightarrow\infty}
\int d\sigma_x dt \left\{
\frac{i\hbar}{2}\left(\psi_{a}^*(\tilde\partial_t\psi_{a})
-(\tilde\partial_t\psi^*_{a})\psi_{a} \right) - \psi_{a}^* H_{a}
\psi_{a}
- \frac{\gamma^{\mu\nu}\overline \epsilon_{{a}{b}}}{4mc^2}
F_{\mu\nu}(\psi^*_{a}\psi_{b} +
\psi_{a}\psi^*_{b}) \right\}
\end{equation}
where the differential operator $H_{a}$ is defined by
\begin{equation}
H_{a} = -\frac{\hbar^2\nabla^2}{2m} + E_{a} + \frac{1}{2mc^2}(E_{a}^2 +
\tilde\partial_t^2),
\end{equation}
and we have redefined the action by a sign in passing to a
Euclideanized non-relativistic metric.  It is now clear that, in the
NR limit $c\rightarrow \infty$, the final two terms in $H_a$ become
negligable, leading to the field equation
\begin{equation}
H_{a}\psi_{a}(x) + {\gamma^{\mu\nu}\overline \epsilon^{{a}{b}} \over 2 m c^2}
F_{\mu\nu}\psi_{b}(x)
= i\hbar\tilde\partial_t \psi_{a}(x),
\label{eq:nr1}
\end{equation}
which is the Sch\"odinger equation of a particle of mass $m$
moving in a constant potential of energy $E_{a}$ with a dipole
interaction.  The dipole interaction term is not
negligeable since the constant $\gamma^{\mu\nu}$ is of order
$c^3$ as we shall show below.

The space-time components $\gamma^{0i}$ can now be related to physical
 electric dipole moments for linear media in the following manner. From
classical electromagnetism we have that the dipole energy
 density is given by ${\bf P}\cdot{\bf E}$, where $\bf P$ is the
 dielectric polarization and $\bf E$ is the electric field. The
 dielectric polarization is related to microscopic displacements of
 charge by
\begin{equation}
{\bf P} = \epsilon_0\chi_e{\bf E} =- \langle e{\bf r}\rangle \times {\rm no.~
density~ of~ charges}.
\end{equation}
If we use the quantum number-density $\psi^*\psi$ here we see that the
dipole energy density is given by
\begin{equation}
{\bf P}\cdot{\bf E} = -\psi^*\psi\;\langle e{\bf r}\rangle\cdot {\bf E}.
\end{equation}
Our non-relativistic Lagrangian is also an energy density, thus
comparing these in the rest frame of the charges, and using the fact
that $F^{0i}= - E^i /c$  we have
\begin{equation}
{\gamma^{\mu\nu} F_{\mu\nu}\over2mc^2}\psi^*\psi =
\frac{\gamma^{0i}}{mc^2}F_{0i}\psi^*\psi =
  - {\gamma^{0i}\over mc^3}E_i \psi^*\psi
=-\langle e{\bf r}\cdot {\bf E}\rangle \psi^*\psi,
\end{equation}
allowing us to identify
\begin{equation}
\gamma^{0i} = mc^3\langle e r^i \rangle.
\label{eq: xx}
\end{equation}
Note that $m$ is the mass of an atom and not the mass of the polarized
charges.  The spatial components $\gamma^{ij}$ are normally zero in
the laboratory frame, but in relatively moving frames they may be
determined by a suitable boost transformation.

\section{Relationship to the Jaynes-Cummings model}

We now wish to show rigorously, making explicit the dimensionless
parameters that have to be small in order for the approximation to
work that, in the non-relativistic limit, our model describes a
two level atomic system interacting via a dipole interaction with an
electromagnetic field. For atoms interacting with a single radiation
mode, the Jaynes-Cummings model emerges naturally.
The Lagrangian for our model is
\be -{\cal L} =
\frac{1}{2}\hbar^2 c^2(\partial_\mu \phi_a)^2 +\frac{1}{2} m_a^2 c^4\phi_a^2
+ \frac{1}{4\mu_0} F_{\mu\nu}F^{\mu\nu} +
\gamma^{\mu\nu}\overline\epsilon_{ab}\phi_a\phi_b\partial_\mu A_\nu.
\label{eq: action}
\ee
which gives for the equation of motion,
\be
\left( -\boxx + \frac{m^2_a c^2}{\hbar^2} \right) \phi_a +
\frac{2}{\hbar^2c^2}\overline\epsilon_{ab}
\phi_b\gamma^{\mu\nu}\partial_{\mu}A_\nu = 0.
\label{eq: m2}
\ee
To take the non-relativistic limit we define two fields\cite{bjorken},
\bea
\psi_a &=& \sqrt{m_a c^2\over2}(\phi_a + \frac{i\hbar}{m_a c^2}
\dot{\phi}_a)\nonumber \\
\chi_a &=& \sqrt{m_a c^2\over2} (\phi_a - \frac{i\hbar }{m_a c^2}
\dot{\phi}_a).
\label{eq: m3}
\eea
The rescaling is necessary in order that the non-relativistic wave-functions
have the right dimensions, with standard inner product.
Clearly, if $\phi_a$ is real, then
 $\psi_a = \chi_a^*$. Moreover, these definitions imply:
\bea
\phi_a &=& {1\over \sqrt{2m_a c^2}}(\psi_a + \chi_a)\\
 i\dot{\phi_a} &=& \left(\frac{m_a c^2}{\hbar}\right){1\over \sqrt{2m_a c^2}}
(\psi_a-\chi_a)
\label{eq: m4}
\eea
from which one can deduce:
\be
i(\dot{\psi}_a + \dot{\chi}_a) = \frac{m_a c^2}{\hbar}(\psi_a - \chi_a)
\label{eq: m5}
\ee
This field redefinition reduces the action to one that is first order
in time derivatives. Using standard
 Legendre transform theory, we obtain the part  of the Hamiltonian density
involving the scalar
fields
\bea
H(\pi_a,\phi_a) &=& \pi_a \dot \phi_a - {\cal L}
\nonumber\\
 &=& \frac{1}{2c^2\hbar^2} \pi_a^2 +\frac{1}{2}\hbar^2 c^2(\nabla \phi_a)^2 +
\frac{1}{2} m^2_a \phi_a^2 +\frac{1}{2} \gamma\cdot F \epsilon_{ab}
\phi_a\phi_b
\eea
where
\bea
\pi_a = c\frac{\delta {\cal L}}{\delta (\dot\phi_a)} = c\hbar^2\dot\phi_a =
-ic\hbar\sqrt{\frac{m_a c^2}{2}}(\psi_a-\chi_a)
\eea
and we have written $\gamma\cdot F \equiv \gamma^{\mu\nu} F_{\mu\nu}$
for brevity.
Replacing $\pi_a$ and $\phi_a$ by their definitions in terms of
$\psi_a$ and $\chi_a$ as given above, we get the Hamiltonian density,
\bea
H(\psi_a,\chi_a)&=& {m_a c^2}\psi_a\chi_a
   +{\hbar\over 4m_a} (\partial_i (\psi_a + \chi_a )
    \partial_i (\psi_a + \chi_a ))
    +{\gamma\cdot F \over 4\sqrt{m_am_b}}\epsilon_{ab}(\psi_a + \chi_a)
     (\psi_b + \chi_b)
\label{eq: hamiltonian density}
\eea
and
\bea {\cal L} &=& \pi_a \dot \phi_a - H(\pi_a,\phi_a) \nonumber \\
&=& {i}\hbar \chi_a\dot{\psi_a}-
   {m_a c^2}\psi_a\chi_a\nonumber\\
  &\ & -{\hbar\over 4m_a} (\partial_i (\psi_a + \chi_a )
    \partial_i (\psi_a + \chi_a ))
    -{\gamma\cdot F \over 4\sqrt{m_am_b}}\epsilon_{ab}(\psi_a + \chi_a)
     (\psi_b + \chi_b)
\label{eq: action2}
\eea
up to total derivatives.
We obtain the equations of motion by varying with respect to $\psi_a$ and
$\chi_a$:
\bea
i\hbar \dot\psi_a &=& -\frac{\hbar^2}{2m_a}\nabla^2(\psi_a + \chi_a) + m_a
c^2 \psi_a + \frac{\epsilon_{ab}\gamma\cdot F}{2c^2\sqrt{m_am_b}}(\psi_b +
\chi_b)
\nonumber \\
i\hbar \dot\chi_a &=& \frac{\hbar^2}{2m_a}\nabla^2(\psi_a + \chi_a) - m_a c^2
\chi_a - \frac{\epsilon_{ab}\gamma\cdot F}{2c^2\sqrt{m_am_b}}(\psi_b +
\chi_b)
\eea
and write the Hamiltonian as a four by four matrix in the 2-component space
$\{a,b\}$ crossed into the 2-component space $\{\psi,\chi\}$:
\bea H_{ab} &=& \left[\delta_{ab}\left(-\frac{\hbar^2\nabla^2}{2m_a} + m_a
c^2\right) + \frac{\gamma \cdot F}{2c^2\sqrt{m_am_b}}\epsilon_{ab}
\right]{\beta}
\nonumber \\
&+& \left[\delta_{ab}\left(-\frac{\hbar^2\nabla^2}{2m_a}\right) + \frac{\gamma
\cdot F}{2c^2\sqrt{m_a m_b}}\epsilon_{ab}\right]{\cal O}
\eea
where
\bea {\beta} =  \left(  \begin{array}{cc}
        1&0\\ 0&-1 \end{array} \right);\,\,\,\,\,\,{\cal O} = \left(
\begin{array}{cc}
        0&1\\ -1&0 \end{array} \right)\\
\eea

So far these equations are exact.
Note that the terms proportional to ${\cal O}$ are non-Hermition and couple
positive and negative
energy states.
We want to perform a  similarity transformation (non-unitary) that will remove
the operator ${\cal O}$ that couples $\psi_a$ and $\chi_a$.  We use an
operator of the form $U_F = e^{i\Lambda}$ where $\Lambda$  has no explicit time
dependence.
We will not be able to find the required $\Lambda$ exactly, so we assume that a
perturbative expansion exists in which $\Lambda$ is small.  In this case,
$\phi ' = e^{i\Lambda}\phi$ and  $\phi ' = H' \phi ' $ which gives,
\bea H' &=& e^{i\Lambda} H e^{-i\Lambda} \nonumber \\
&=&   H +i[\Lambda,H] +\,\,.\,\,.\,\,.\,\,
\eea
We will in fact need three independent expansion parameters.
 In addition to the usual  non-relativistic expansion parameters $\lambda_a$
for the two atomic states:
\be
\lambda_a = {\hbar^2 \nabla^2\over  m_a^2 c^2}
\ee
we will need  the coupling expansion parameter:
\be
\lambda_3 = {\gamma\cdot F \over \overline m c^2 \sqrt{m_1m_2}c^2}
\ee
where $\overline m=\frac{1}{2}(m_1+m_2)$. We will also assume that
$(m_1-m_2)<<\overline m$. Since, as argued above,
$\gamma\cdot F \propto mc^2 e \langle {\bf r} \cdot {\bf E}\rangle$, it follows
that
\be
\lambda_3\sim  {e \langle {\bf r} \cdot {\bf E}\rangle \over
\overline m c^2}
\ee
 Thus $\lambda_3 <<1$ requires
the dipole energy in the electric field to be much smaller than the rest energy
of the atom.

 We can now expand $\Lambda$:
\be
\Lambda = \Lambda_1 + \Lambda_2 + O(\lambda^2)
\ee
where $O(\lambda^2)$ refers to a product of any two of the small expansion
parameters, and
\bea
\Lambda_1 &=& \frac{i}{2} \lambda_a \delta_{ab}\beta {\cal O}\\
\Lambda_2 &=& -\frac{i}{2} \lambda_3 \epsilon_{ab} \beta {\cal O}
\eea
In the above
\bea \beta{\cal O} =  \left(  \begin{array}{cc}
        0&1\\ 1&0 \end{array} \right).
\eea
It is easy to verify that $i[\Lambda,H]$ to leading order in $\lambda$ exactly
cancels the terms in $H$ that are proportional to ${\cal O}$.  This decoupling
of the $\psi$ and $\chi$ modes yields a Hamiltonian which can be written
as a two by two matrix in the space \{a,b\} acting on the column vector
\bea { \psi_1 \choose \psi_2}. \nonumber \eea
In particular (dropping the prime):
\be
H = H_{0} + H_{int}
\ee
where the free part of the Hamiltonian is:
\bea
H_{0}= \left(
\begin{array}{cc}
-\frac{\hbar^2 \nabla^2}{2m_1} + m_1 c^2 & 0 \\
0 & -\frac{\hbar^2 \nabla^2}{2m_2}+m_2 c^2
\end{array}
\right).
\eea
This result is almost the same as Korenman's\cite{korenman1}, and
differs only by the fact that the kinetic terms have different
masses---a consequence of the fact that we have chosen to view the
shifted masses $m_a$ as fundamental.  As seen below, this only gives a
higher order correction which can be neglected in the non-relativistic
limit.  The interaction Hamiltonian is:
\bea
H_{int} =
\left(
\begin{array}{cc}
0 & \frac{\gamma\cdot F}{2c^2\sqrt{m_1m_2}} \\
\frac{\gamma\cdot F}{2c^2\sqrt{m_1m_2}} & 0
\end{array}
\right),
\eea
After some algebra the Hamiltonian can be written in the following
form:
\bea
H &=& \left( -\frac{\hbar^2\nabla^2}{2 \overline m} + \overline m c^2\right)
\left(
   \begin{array}{cc}1&0\\  0&1
     \end{array}\right)
 +  {\hbar\omega_{12}\over2}  \left(  \begin{array}{cc}
        1&0\\ 0&-1 \end{array} \right)\nonumber\\
 & & \, +{\gamma\cdot F\over 2c^2\sqrt{m_1m_2}} \left(\begin{array}{cc}
      0&1\\1&0
   \end{array} \right) + {\hbar\omega_{12}\over 2} \left(
   \begin{array}{cc}
     {\hbar^2\nabla^2\over 2\overline m m_1 c^4}&0  \\
     0 & {-\hbar^2\nabla^2 \over 2 \overline m m_2 c^4}
 \end{array}\right)
\label{eq: free hamiltonian}
\eea
where we have defined the energy difference
$\omega_{12} = (m_1 - m_2)c^2/\hbar$.

The first three terms have a very natural physical interpretation: The
first term is the free Hamiltonian for the ``collective modes'' of the
atoms (the term proportional to the mass is just a shift in the energy
and not relevant), while the next two describe the energy splitting
and the corresponding dipole interaction with the electromagnetic
field.  Assuming that $\hbar\omega_{12}<<\overline mc^2$, the last
term is an order $\lambda$ correction to the second term. It is
therefore higher order in the non-relativistic expansion and
consistency demands that we neglect it.

We drop the term in the Hamiltonian that corresponds to the collective modes
and write the remaining piece as the sum of two terms:
\bea
H = \frac{1}{2}\hbar\omega_{12} \sigma_z + \frac{\gamma\cdot
F}{2c^2\sqrt{m_1m_2}}(\sigma_+ + \sigma_-) \eea
where $\sigma_z$ is the third Pauli matrix and $\sigma_+$ and $\sigma_-$ are
the usual raising and lowering operators for the atomic states:
\bea
\sigma_+ &=& \left(\begin{array}{cc}
     0&1\\ 0&0
    \end{array}\right) \\
\sigma_- &=&
 \left(\begin{array}{cc}
     0&0\\ 1&0
    \end{array}\right)
\eea
We can now make contact with the Jaynes-Cummings model by assuming a
single mode electric field, linearly polarized in the $x$-direction,
as would be found in a high-Q cavity of volume $V$, for example. In
terms of the standard harmonic oscillator creation and annihilation
operators, the field can be written\cite{meystre_sargent}:
\be
\vec{E} = \hat{x}{\cal E}_\Omega (a + a^\dagger) \sin Kz
\ee
%
%
where
${\cal E}_\Omega = [\hbar\Omega/\epsilon_0 V]^{1\over2}$ is the ``electric
field per
photon'' for an electric field of frequency $\Omega$. In the above, $\hat{z}$
points along the
longitudinal axis of the cavity and
$K = \Omega /c$ is the magnitude of the corresponding wave number.

The interaction Hamiltonian now takes the form
\bea
H_{int} &=& -{\gamma_{0i} E^{i} \over  c^3 \sqrt{ m_1m_2}}
     (\sigma_+ + \sigma_-)(a + a^\dagger)\\
   &=& - {\gamma_x {\cal E}_\Omega\over c^3 \sqrt{m_1m_2} }\sin Kz
(\sigma_+ + \sigma_-)(a + a^\dagger)
\eea
where $\gamma_x$ is the component of $\gamma_{0i}$ in the direction of the
electric field.

We can drop terms proportional to $\sigma_- a$ and $\sigma_+ a^\dagger$. These
terms correspond to the simultaneous lowering of an atom and absorption of a
photon, and the simultaneous raising of an atom and production of a photon, and
we expect them to be suppressed.  We can see that this is the case by looking
at the evolution of the operators in the Heisenberg picture.  Writing
\bea
\sigma_{\pm}(t) &=& \sigma_{\pm}(0) e^{\pm i\omega t} \nonumber \\
a(t) &=& a(0) e^{-i\Omega t} \nonumber \\
a^\dagger(t) &=& a^\dagger(0)e^{i\Omega t} \nonumber
\eea
we find that $\sigma_- a$ and
$\sigma_+ a^\dagger$ are proportional to  $e^{\pm i(\omega+\Omega)t}$ and the
other two products are proportional to $e^{\pm i(\omega-\Omega)t}$.  We are
interested in a system that is tuned close to resonance $\Omega \approx
\omega$ and therefore, in the random phase approximation, terms proportional to
$e^{\pm i(\omega+\Omega)t}$ will average to zero because of the rapid
oscillation of the phase.
The final  result has precisely the form of the interaction term for
the Jaynes-Cummings model:
\bea
H_{int} = \hbar g (a \sigma_+ + a^\dagger \sigma_-)
\eea
and we identify the Rabi frequency in our model as:
\be
g \equiv -{\gamma_x {\cal E}_\Omega {\rm sin}Kz \over \hbar c^3 \sqrt{m_1 m_2}
}
\ee
This corresponds to the usual Rabi frequency\cite{meystre_sargent}
\be
g_R = - {\langle e \vec{x}\rangle_x {\cal E}_\Omega \over  \hbar}\sin Kz
\ee
on making the identification:
\be
\gamma_x = { \langle e \vec{x} \rangle_x} \sqrt{m_1 m_2} c^3
\ee
which is consistent with the identification for $\gamma$ made in the
previous section (eqn. (\ref{eq: xx})), apart from  terms of order
$(m_1-m_2) / m$.

\section{The effective action and its interpretation}

Having established a connection to the Jaynes-Cummings model, we no
longer need to refer to it and we can focus entirely on the
relativistic case. Quantum corrections to the relativistic model may
be computed using standard field theoretical prescriptions. The
effective action is a particularly elegant way of generating such
corrections.  Although our theory is already an effective theory, this
does not invalidate the procedure of looking for corrections due to
correlations in our chosen field variables. If such corrections were
already accounted for, they would simply renormalize away trivially in
a renormalizable theory. In a non-renormalizable theory, this is not
possible by definition and thus the corrections are always of
interest. The calculation of loop corrections in non-renormalizable
field theories has been shown to give accurate physics. See
ref. \cite{weinberg1} for a discussion of this.

The effective action is a generating functional for
one-particle irreducible quantum corrections. It
describes completely the behaviour of the averaged
field (or background field) after all quantum fluctuations
have been taken into account. If one defines the functional
integral for a field $\Phi$ by
\begin{equation}
W[J] = -i\ln \int d\mu[\Phi] \exp\left\{ iS[\Phi] + \int dV\; J\Phi\right\}
\label{eq:28}
\end{equation}
where $\hbar=c=\mu_0=\epsilon_0=1$,
and the c-number average field by
\begin{equation}
\overline \Phi = \langle\Phi\rangle = \frac{\delta W}{\delta J}
\label{eq:29}
\end{equation}
then the effective action is given by the Legendre transform
of $W[J]$ which displaces the explicit dependence on the source of
fluctuations $J$, in favour of a dependence on the average field itself.
\begin{equation}
\Gamma[\overline\Phi] = W[J] - \int dV\; J\overline \Phi
\label{eq:30}
\end{equation}
The resulting object is in all senses an action for the
average field. In the remainder of the paper we make use
of the background field method to compute the effective
action. We begin by dividing the field into an average
part and a fluctuating part for convenience:
\begin{equation}
\phi_{a} \sim \overline \phi_{a} + \varphi_{a},
\label{eq:31}
\end{equation}
where $\overline \phi_{a}$ is the average field and $\varphi_{a}$ is
the quantum field which replaces the total $\phi$ as the variable of
integration in eqn (\ref{eq:28}).  This division may now be used as a
basis for generating a perturbation expansion for the effective
action.  Our model for the laser contains two fields $\phi_{a}$ and
$A_\mu$. We shall assume that the average external field
$F_{\mu\nu}=0$, so that $\overline A_\mu = \frac{\delta W}{\delta
J_\mu}$ may always be gauged to zero in all physical results, provided
only that the systems lives in a box with a simple topology.  It is
nevertheless useful to keep this quantity non-zero when using
$\Gamma[\overline\phi,\overline A_\mu]$ as a generating functional,
since derivatives with respect to the vector field allow us to easily
calculate physical quantities of interest.  We expand the action
around these background fields
\begin{equation}
S\left[\overline \phi_{a} + \varphi_{a}, \overline A_\mu + A_\mu \right]
= S_{\rm class}[\overline \phi_a,\overline A_\mu] + S_2[\varphi_{a},A_\mu]
+ S_{\rm int}[\overline\phi_{a},\varphi_{a}, A_\mu],
\label{eq:32}
\end{equation}
where $S_{\rm class}$ is the term composed purely of background
fields, $S_2$ is quadratic in the quantum field variables
and $S_{\rm int}$ is the remainder. The effective action is then
given by the one-particle irreducible part of
\begin{eqnarray}
\Gamma[\overline \phi_{a}, \overline A_\mu] &=& S[\overline\phi_{a}, \overline
A_\mu] -i \ln \int
d\mu[\varphi_{a},A_\mu]
e^{iS_{\rm int}}e^{iS_2}\nonumber\\
&=& S[\overline\phi_{a}, \overline A_\mu] + \langle S_{\rm int} \rangle + i
\langle
(S_{\rm int})^2\rangle +\ldots
\label{eq:33}
\end{eqnarray}
We compute the effective action in two stages. First we
consider fluctuatations in the photon field
leading to an intermediary effective action $\Gamma_A$. These can be
dealt with exactly and this leaves us with a result for
the dynamics of the atomic system with all photon
degrees of freedom eliminated. This is action could
then be used to describe the situation in the micro-cavity
maser where the measurable degree of freedom are the
atomic states, and the effects of the photons are
only felt indirectly. Secondly, we consider fluctuations in
the atomic degrees of freedom, such as one would expect
in a gaseous or solid state laser. This gives us the
full effective action $\Gamma[\overline\phi_{a}]$.

We begin by considering the radiation field with
a Lorentz gauge fixing term added and associated Lagrange
multiplier ${1/\alpha}$,
\be
S_M = \int dV_x \left\{ \frac{1}{4}F^{\mu\nu}F_{\mu\nu} +
P^{\mu\nu}F_{\mu\nu}
+\frac{1}{2{\alpha}}(\partial_\mu A^\mu)^2 \right\}.
\label{eq:34}
\ee
Ghost terms may be absorbed into the functional measure in view
of the trivial nature of the gauge field contribution.

The functional integral over $A_\mu$ may be performed immediately
since it is Gaussian.
Integrating by parts and shifting the quantum gauge field (the field
of integration) $A_\mu \rightarrow A_\mu - 2 \partial^\nu P_{\mu\nu}$, one
obtains
without modification to the functional measure,
\begin{equation}
S_M =  \int dV_x \left\{
\frac{1}{2}A^\mu\left[-\boxx\; \delta_\mu^{~\nu} +
\left(1-\frac{1}{{\alpha}}\right)\partial_\mu\partial^\nu \right]
A_\nu - 2 \int dV_{x'}(\partial_\mu P^{\mu\nu})D_{\nu\sigma}(\partial_\rho
P^{\rho\sigma})
\right\}
\label{eq:35}
\end{equation}
The integral over the gauge field $A_\mu$ is now a Gaussian and may be
dealt with by standard results. This results only in a
constant addition to the effective action which may be renormalized
away by a shift of the arbitrary zero point for the energy scale.  The
result is the one-loop correction
\begin{equation}
\Gamma_A^{(1)}[P^{\mu\nu}] = {\rm const} + 2\int dV_x
dV_{x'}\;(\partial_\mu P^{\mu\nu})D_{\nu\sigma}(\partial_\rho P^{\rho\sigma})
\label{eq:36}
\end{equation}
where the free photon Green function is defined by the relation
\begin{equation}
\left[-\boxx\; g_{\mu\nu} +
\left(1-\frac{1}{{\alpha}}\right)\partial_\mu\partial_\nu \right]
D^{\nu\rho} = \delta_\mu^{~\rho}\delta(x,x').
\label{eq:37}
\end{equation}
Using this general result we obtain the first stage effective action
for atomic field $\phi_{a}$:

\begin{eqnarray}
\Gamma_A[\phi] = \int dV_x
\left\{
\frac{1}{2}\phi_{a}\left[-\boxx +m_{a}^2\right]\phi_{a}
+ 2 \int dV_{x'}
\phi_{a}(x)\phi_{b}(x)\overline
V^{{a}{b}{c}{d}}(x,x')\phi_{c}(x')\phi_{d}(x')
\right\}
\label{eq:40}
\end{eqnarray}
and
\begin{equation}
\overline V^{{a}{b}{c}{d}}(x,x') =
\gamma^{\mu\nu}_{{a}{b}}\gamma^{\rho\sigma}_{{c}{d}}
(\stackrel{x}{\partial}_\mu \stackrel{x'}{\partial}_\rho
D_{\nu\sigma}(x,x')).
\label{eq:41}
\end{equation}
where we have introduced the short hand notation $\gamma^{\mu\nu}_{{a}{b}}
=\gamma^{\mu\nu}\overline\epsilon_{{a}{b}}$.

To generate the second stage effective action, we expand the atomic variables
about
a background or external field. The physical significance of this step is
the presence of measurable averages for the atomic variables in our system.

Expanding around free fields and dealing with the interaction term as
an expansion of the exponentiated action, we get,
\begin{eqnarray}
\Gamma[\overline\phi] &=& \Gamma_A[\overline\phi_{a}]  -i \ln \int
d\mu[\varphi_{a}] \left(e^{iS_{\rm int}}\right)
\exp\left\{iS^{(2)}\right\}\label{eq:42}\\
&=& \langle S_{\rm int}\rangle + i \langle (S_{\rm int})^2\rangle + \ldots
O(\overline \phi^3)
\label{eq:43}
\end{eqnarray}
The Feynman (time-ordered) propagator is defined by
\begin{equation}
\langle \varphi_{a}(x) \varphi_{b}(x') \rangle = -i
G_{{a}{b}}\delta(x,x').
\label{eq:44}
\end{equation}

For the renormalizable $\overline V$ vertex we now obtain the
part of the effective action which
is quadratic in the background fields. The outstanding terms do not contribute
to the
self energy and therefore to the decay rates of the atomic levels.
\begin{eqnarray}
\langle S_{\rm int} \rangle &=& -i \int dV_x dV_{x'} \overline
V^{{a}{b}{c}{d}}(x,x')
\Big\langle
\overline \phi_{a}(x)\overline\phi_{d}(x')G_{{b}{c}}(x,x')
+ \overline\phi_{a}(x)
\overline\phi_{c}(x')G_{{b}{d}}(x,x')\nonumber\\
&+& \overline\phi_{b}(x)\overline\phi_{c}(x')G_{{a}{d}}(x,x')
+ \overline\phi_{b}(x)\overline\phi_{d}(x')
G_{{a}{c}}(x,x')\Big\rangle + {\rm disconnected}
\label{eq:47}
\end{eqnarray}
The matrix $\gamma_{{a}{b}}^{\mu\nu}$ is off-diagonal but symmetrical
in $a,b$, so
we may write
\begin{equation}
\overline V^{{a}{b}{c}{d}} = \left\{
\overline V^{1212} = \overline V^{2112} = \overline V^{2121} = \overline
V^{1221} = \gamma^{\mu\nu}\gamma^{\rho\sigma}
(\stackrel{x}{\partial}_\mu \stackrel{x'}{\partial}_\rho D_{\nu\sigma}(x,x'))
\right\}.
\label{eq:48}
\end{equation}
This last result gives us
\begin{eqnarray}
\langle S_{\rm int}\rangle &=& -4i\gamma^{\mu\nu}\gamma^{\rho\sigma} \int dV_x
\int dV_{x'}
(\stackrel{x}{\partial}_\mu \stackrel{x'}{\partial}_\rho
D_{\nu\sigma}(x,x'))\nonumber\\
&~& \times \left\{ \overline\phi_1(x)\overline\phi_1(x') G_{22}(x,x')
+\overline\phi_2(x)\overline\phi_2(x')G_{11}(x,x')\right\}.
\label{eq:49}
\end{eqnarray}
We are interested in the effective coupling constants of the quantized
theory, which may be defined through derivatives of the effective
action.  These provide us with information about the decay rates or
lifetimes of the atomic levels and corrections to the Rabi-flopping
frequency.  The momentum-space structure of these quantites also
illustrate how photon energies are related to the interatomic spacings
etc.  Specifically, we wish to compute the diagonal scalar self-energy
$\Sigma_{aa}$, whose imaginary part gives an indication of the decay
rates of the levels,
\begin{equation}
\Sigma_{aa}(x,x') =
\frac{\delta^2\Gamma[\overline \phi]}
{\delta\overline\phi_a(x)\delta\overline\phi_a(x')},
\end{equation}
the interaction vertex (or generalized coupling constant)
\begin{equation}
\Gamma_\mu(x,x',x'') = \overline
\epsilon_{ab}\frac{\delta^2\Gamma[\overline \phi,\overline A_\lambda]}
{\delta\overline\phi_a(x)\delta\overline\phi_b(x')\delta \overline A^\mu(x'')}
\end{equation}
and the photon self-energy or polarization tensor
\begin{equation}
\Pi_{\mu\nu}(x,x') = \frac{\delta^2\Gamma[\overline \phi, \overline A_\lambda]}
{\delta \overline A^\mu(x)\delta \overline A^\nu(x')}.
\end{equation}

\section{One-Loop Renormalizability}
We now verify the one-loop renormalizability of our theory.  For the
remainder of the paper, we choose natural units in which
$\hbar=c=\epsilon_0=\mu_0=1$.  General arguments indicate that
renormalizablity is connected with power-counting, or the dimension of
the coupling constant.  In this scheme there is only one scale of
dimensions.  Length and time are completely equivalent and mass
is the inverse of length. A dimensional analysis of the action in
these units leads to the conclusion that both the scalar and vector
field $\phi_a$ and $A_\mu$ in $n+1$ space-time dimensions has
engineering dimension
\begin{equation}
[\phi_a] = [A_\mu] = L^{\frac{1-n}{2}}.
\end{equation}
A renormalizable quantum field theory is one in which all the
infinities accrued by the calculational procedure can be defined away
by reinterpreting the coupling constants appearing in the action. This
is possible only if the infinite terms are of the same form as the
original terms in the action which contain the coupling constants. In
a non-renormalizable theory, it is not possible to absorb all infinities with a
finite number of redefinitions.

For the moment we will consider both the
non-renormalizable and renormalizable interactions, in order to contrast
them:
\begin{eqnarray}
\tilde P_{\mu\nu}: ~~~ [\gamma^{\mu\nu}] &=& L^{\frac{n-1}{2}}\nonumber\\
 P_{\mu\nu}: ~~~ [\gamma^{\mu\nu}] &=& L^{\frac{n-3}{2}}
\label{eq: yy}
\end{eqnarray}
We may consider the case of both two and three spatial
dimensions, since a laser often has an axial symmetry which reduces
its effective dimensionality.  For the first of the interactions in eqn.
(\ref{eq: yy}) $\gamma^{\mu\nu}$
has the dimensions of $L$ in $3+1$ dimensions and $L^{\frac{1}{2}}$ in
$2+1$ dimensions.  In both cases the interaction is non-renormalizable.  The
second
interaction is more successful. In $3+1$ dimensions,
$\gamma^{\mu\nu}$ is dimensionless which implies that
the theory is strictly (also called marginally) renormalizable. In
$2+1$ dimensions, $\gamma^{\mu\nu}$ has the dimensions
of $L^{-\frac{1}{2}}$, which implies that the theory is
super-renormalizable.

Using the second interaction, the Jaynes-Cummings model can be represented as a
renormalizable field theory given
by the action in eqns. (\ref{eq:2}) and (\ref{eq:12}).  We note that, although
renormalizablity is often
regarded as a critereon for choosing between field theories, it is not
an infallible guide to their physicality. Quantum corrections to
non-renormalizable theories are known to give accurate results in a
number of cases\cite{weinberg1}. Moreover, we have a natural energy
cut-off for the kinetic motion of atoms, namely $kT$.  Our primary
reason for choosing the renormalizable interaction is that it is
easier to calculate quantum corrections in this case; the lack
of an explicit time-derivative preserves Lorentz covariance.

\subsection{Scalar self-energy}

We seek to calculate the one-loop self energy $\Sigma(p)$ and vertex
function $\Gamma_{\mu}(p,p',q)$ and show that these terms have
infinite pieces that have the same form as the original interaction,
and thus can be reabsorbed into the coupling constants, the masses,
and rescaling factors.  We use cutoff regularization since we
ultimately want to use our model to study laser physics, which will
involve the imposition of boundary conditions.  We define the
subtraction scheme by expanding around the mass shell.  We expand in
$\frac{\delta}{M^2}$ where $\delta = m_1^2-m_2^2$ and
$M^2=\frac{1}{2}(m_1^2+m_2^2)$.

The bare theory gives a propagator of the form,
\bea iG_{aa}(p) = \frac{i}{p^2 + m_{a0}^2 - i\epsilon}
\nonumber
\eea
which has a pole at $p^2 = -m^2_{a0}$.  We calculate the polarization tensor
and use the Dyson equation to obtain a propagator of the form,
\bea
 iG_{aa}(p) = \frac{i}{p^2+ m_{a0}^2 + \Sigma_{aa}(p) - i\epsilon} \nonumber
\eea
We define $\Delta m_a^2 = m_a^2-m_{a0}^2$ and write,
\bea
iG_{aa}(p) = \frac{i}{p^2 + m_a^2 +[\Sigma_{aa}(p) -\Delta
m_a^2]-i\epsilon}\nonumber
\eea
We choose
\bea
\Delta m_a^2 = \Sigma_{aa}(- m_a^2)
\label{1}
\eea
so that the pole occurs at $p^2=-m_a^2$ which we call the physical mass.  Thus,
the propagator can be written,
\bea
iG_{aa}(p) = \frac{i}{p^2 + m_a^2 + [\Sigma_{aa}(p^2) - \Sigma_{aa}(-m^2_a)] -
i\epsilon}\nonumber
\eea
$\Sigma_{aa}(p)$ is divergent and a requirement for renormalizabilty is that we
can write (after regularization)
\bea
\Sigma_{aa}(p^2) - \Sigma_{aa}(-m_a^2) = (p^2 + m_a^2) f(\Lambda)
\label{2}
\eea
so that the propagator becomes
\bea
iG_{aa}(p) = \frac{iZ_{\phi_a}}{p^2+m_a^2-i\epsilon};\,\,\,\,\,Z^{-1}_{\phi_a}
=
1+f(\Lambda)
\label{3}
\eea
These redefinitions are equivalent to the statement that we can add
counterterms to the Lagrangian of the form,
\bea
{\cal L}^1_{ct} = \frac{1}{2}(Z_{\phi_a}^{-1} -1)((\partial_\mu \phi_a)^2 +
m_a^2 \phi_a^2) + \frac{1}{2} \Delta m_a^2 \phi_a^2 \nonumber
\eea and absorb the infinities from the self energy $\Sigma_{aa}(p)$ in the
mass
shift $\Delta m_a^2$ and the wavefunction scaling factor $Z_{\phi_a}$.

In a similar way, the interaction part of the Lagrangian  ${\cal L}_{int}$
gives
rise to a bare vertex of the form,
\bea
\Gamma^{(0)}_\mu = 2\epsilon_{ab}q_{\mu'}\gamma_0^{\mu'\mu} \nonumber
\eea
where $q$ is the incoming photon momentum and $\gamma_0^{\mu\mu'}$ is the bare
coupling constant.  The one loop contribution to this vertex is divergent.  We
isolate the divergent part by performing a subtraction at the mass shell,
\bea
\Gamma_\mu^{(1)}(p,p',q) = \Gamma_\mu^{(1)}\Big|_{ms} +
\tilde{\Gamma}_\mu^{(1)}
\nonumber
\eea
where the subscript $ms$ means that the external momenta are evaluated on the
mass shell, and $\tilde{\Gamma}$ is finite. A requirement of renormalizability
is that we can write, after regularization,
\bea
\Gamma_\mu^{(1)}\Big|_{ms} = \Gamma_\mu^{(0)} (Z_1^{-1} -1) \label{4}
\eea
which means that we can absorb the infinite part of the one loop vertex graph
into a redefinition of the coupling constant.  This redefinition is equivalent
to adding to the Lagrangian a counterterm of the form
\bea
{\cal L}^{2}_{ct} = (Z_1^{-1} -1) {\cal L}_{int} \nonumber \eea

In this section, we will calculate the one loop self energy and the one loop
vertex function and use~(\ref{1}),~(\ref{2}),~(\ref{3}) and~(\ref{4}), to
determine $\Delta m_a^2$, $Z_{\phi_a}$ and $Z_1$.  We start from the following
expression for the self-energy of the scalar field $\phi_1$:
\bea
\Sigma_{11} = -4i\gamma^{\tau\mu}\gamma^\lambda_{\,\,\mu} \int \frac{d^4
k}{(2\pi)^4}
\frac{k_\tau k_\lambda}{(k^2 - i\epsilon)((k+p)^2 + m_2^2 - i\epsilon)}
\nonumber
\eea
where we have used the Feynman gauge $\alpha=1$ for the internal photon
propagator.  The
self-energy for the field $\phi_2$ will depend on $m_1$ in the same way.  We
rewrite the denominator using the usual Feynman parameter formula,
\bea
\frac{1}{k^2- i\epsilon}\frac{1}{(k+p)^2 + m_2^2 - i\epsilon} = \int _0^1 dx
\frac{1}{(k^2(1-x) + [(k+p)^2 + m_2^2]x - i\epsilon)^2}
\nonumber
\eea
We complete the square in the denominator and shift the integration variable
$k=l-px$ to obtain,
\bea
\Sigma_{11} = -4i\gamma^{\tau\mu}\gamma^\lambda_{\,\,\mu} \int \frac{d^4
l}{(2\pi)^4}
\int_0^1 dx \frac{l_\tau l_\lambda + x^2p_\tau p_\lambda}{(l^2 + a^2 -
i\epsilon)^2}
\nonumber
\eea
where $a^2 = m_2^2x + p^2x(1-x)$ and we have dropped the terms linear in $l$
which give zero by symmetric integration.  We do a Wick rotation so that the
integration contour lies along the imaginary axis and make the change of
variable,
$l_0 = il_4$ to obtain the Eucledian space integral,
\bea
\label{5}
\Sigma_{11} = 4\gamma^{\tau\mu}\gamma^\lambda{\,\,_\mu} \int \frac{d^4
l}{(2\pi)^4}
\int_0^1 dx \frac{l_\tau l_\lambda + x^2 p_\tau p_\lambda}{(l^2 + a^2 )^2}
\eea
The integral is infinite and we use cut-off regularization to render it finite.
 After regularization we can switch the order of integration and perform the
$l$ integration first.  We consider the two pieces separately.  First we
evaluate the term proportional to $\l_\tau l_\lambda$ and call it
$\Sigma_{11}^I$.  Under the integral sign we can replace $l_\tau l_\lambda$ by
$\frac{1}{4}g_{\tau\lambda}l^2$ (by symmetric integration) which gives,
\bea
\Sigma^I_{11} = \gamma^2\int_0^1 dx   \int \frac{d^4
l}{(2\pi)^4}\frac{l^2}{(l^2+a^2)^2}
\nonumber
\eea
where $\gamma^2 = \gamma_{\mu\nu}\gamma^{\mu\nu}$.
Doing the $l$ integration gives,
\bea
\Sigma^I_{11} =
\frac{\gamma^2}{(2\pi)^3}\int_0^1 dx (\Lambda^2+a^2-2a^2{\rm ln}
\frac{\Lambda^2}{a^2})
\nonumber
\eea
We expand in ${b}=\delta/M^2$ and take only the leading order term.  We
calculate $\Sigma^I_{11}(-m_1^2)$ and $
\Sigma^I_{11}(p^2)-\Sigma^I_{11}(-m_1^2)$.
The result is,
\bea \Sigma^I_{11}(-m_1^2) &=& \frac{\gamma^2}{(2\pi)^3}(\Lambda^2 -
\frac{2}{3}M^2 {\rm ln}\frac{\Lambda^2}{M^2} - \frac{M^2}{9})\nonumber\\
\Sigma^I_{11}(p^2)-\Sigma^I_{11}(-m_1^2) &=&
-\frac{\gamma^2}{(2\pi)^3}[\frac{25}{18} (p^2 + m_1^2) + \frac{1}{3}(p^2 +
m_1^2){\rm ln}\frac{\Lambda^2}{M^2}]
\label{6}
\eea
Next we have to calculate the term proportional to $p_\alpha p_{\beta}$.  From
{}~(\ref{5}) we have,
\bea
\Sigma_{11}^{II} = 4\gamma^2_{\tau\lambda}q^\lambda q^\tau \int^1_0 dx\,\,x^2
\int \frac{d^4l}{(2\pi)^4} \frac{1}{l^2+a^2}
\nonumber
\eea
where $\gamma^2_{\tau\lambda} = \gamma^{\mu}_{\,\,\tau}\gamma_{\mu\lambda}$.
Doing the $l$ integration we obtain,
\bea \Sigma_{11}^{II}(p) = \frac{4\gamma^2_{\lambda\tau}}{(2\pi)^3} p^\tau
p^\lambda \int _0^1 dx\,\,x^2 ({\rm ln}\frac{\Lambda^2}{a^2}-1)
\nonumber
\eea
which gives,
\bea
\Sigma^{II}_{11}(-m_1^2) &=& \frac{4\gamma^2_{\tau\lambda}}{(2\pi)^3} p^\lambda
p^\tau[\frac{1}{3}{\rm ln}\frac{\Lambda^2}{M^2} - \frac{1}{9}] \nonumber \\
\Sigma^{II}_{11}(p^2) - \Sigma^{II}_{11}(-m_1^2) &=&
-\frac{2\gamma^2_{\tau\lambda}}{3(2\pi)^3}p^\tau p^\lambda [\frac{p^2 +
m_1^2}{M^2}]
\label{7}
\eea
Thus, from~(\ref{1}),~(\ref{2}),~(\ref{3}),~(\ref{6}) and~(\ref{7}) we obtain,
\bea
\Delta m_a^2 &=& \frac{\gamma^2}{(2\pi)^3}M^2 [\frac{\Lambda^2}{M^2}
-\frac{2}{3}{\rm ln}\frac{\Lambda^2}{M^2} - \frac{1}{9} +
4\frac{\gamma^2_{\tau\lambda}p^\tau p^\lambda}{\gamma^2 M^2}[\frac{1}{3}{\rm
ln}\frac{\Lambda^2}{M^2} - \frac{1}{9}]]\\
Z_{\phi_1}^{(-1)} &=& 1-\frac{\gamma^2}{(2\pi)^3}[\frac{25}{18} +
\frac{1}{3}{\rm ln}\frac{\Lambda^2}{M^2} +
\frac{2}{3}\frac{\gamma^2_{\tau\lambda}p^\tau p^\lambda}{\gamma^2 M^2}]
\label{8}
\eea
and $Z_{\phi_2} = Z_{\phi_1}$.  Keeping only the divergent terms,
$Z_{\phi_a}^{(-1)}$ is a wavefunction renormalization factor of the usual form,
and the first two terms in $\Delta m_a^2$ give an infinite shift in the mass
term in the standard way.  The fourth term in the expression for $\Delta m_a^2$
corresponds to a new interaction in the Lagrangian at the one loop level of the
form,
\bea
\gamma^2_{\mu\nu}(\partial^\mu \phi)(\partial^\nu \phi) \nonumber
\eea

\subsection{Interaction vertex}

Next we obtain the vertex renormalization constant from the one loop vertex
correction shown in Fig XX.  We obtain,
\bea
\Gamma_\mu^{(1)} = 8i\gamma^\alpha_{\,\,\lambda
}\gamma^{{b}\lambda}\gamma^{\mu '\mu}q_{\mu '} \int \frac{d^4 k}{(2\pi)^4}
\frac{k_\alpha k_{b}}{(k^2-i\epsilon)((p'-k)^2+m_1^2-i\epsilon)
((p-k)^2+m_2^2-i\epsilon)}
\nonumber
\eea
We rewrite the integral in terms of two Feynman parameters by using the
expression,
\bea
\frac{1}{ABC} = 2\int_0^1 x\,\,dx\int_0^1 dy \frac{1}{(xyA + x(1-y)B +
(1-x)C)^3}
\nonumber
\eea
with
\bea
A&=& (p-k)^2 + m_2^2 - i\epsilon \nonumber\\
B&=& (p'-k)^2 + m_1^2 - i\epsilon \nonumber\\
C&=& k^2 - i\epsilon.
\nonumber
\eea
Putting the external scalars on the mass shell we obtain,
\bea
\Gamma_\mu^{(1)} = 16i\gamma^\alpha_{\,\,\lambda}
\gamma^{\beta\lambda}\gamma^{\mu '\mu}q_{\mu '} \int \frac{d^4 k}{(2\pi)^4}
\int_0^1x\,\,dx \int_0^1 dy \frac{k_\alpha k_\beta}{([k-(pxy+p'x(1-y)]^2 + b^2
- i\epsilon)^3}
\nonumber
\eea
where
\bea
b^2 = x^2m_2^2 + \delta x^2(1-y) + x^2y(1-y)q^2
\nonumber \eea
We shift the integration variable
\bea
l = k-(pxy+p'x(1-y))
\nonumber
\eea and drop the terms linear in $l$ which give zero by symmetric integration.
 We perform a Wick rotation so that the integration contour lies along the
imaginary axis, and make the change of variable $k_0=ik_4$.  The result is,
\bea
\Gamma_\mu^{(1)} = -16\gamma^\alpha_{\,\,\lambda}
\gamma^{\beta\lambda}\gamma^{\mu '\mu}q_{\mu '} \int \frac{d^4 l}{(2\pi)^4}
\int_0^1x\,\,dx \int_0^1 dy \frac{l_\alpha l_\beta +
M_{\alpha\beta}}{(l^2+b^2)^3}
\nonumber
\eea
where,
\bea
M_{\alpha\beta} = (p_\alpha xy + p'_\alpha x(1-y))(p_\beta xy + p'_\beta
x(1-y))
\nonumber
\eea

We consider separately the terms proportional to $l_\alpha l_\beta$ and
$M_{\alpha\beta}$.  We will first do the integral for the term containing
$l_\alpha l_\beta$ and call it $\Gamma_\mu^{(1)I}$.  This term is divergent,
and we use cutoff regularization.  By symmetric integration we can write
$l_\alpha l_\beta = \frac{1}{4} g_{\alpha\beta}l^2$ under the integral sign.
Switching the order of integration and performing the $l$ integration gives,
\bea
\Gamma_\mu^{(1)I} = -4 \gamma^2 \gamma^{\mu ' \mu}q_{\mu '} \int_0^1 x\,\,dx
\int_0^1 dy ({\rm ln}\frac{\Lambda}{b} - \frac{3}{4}).
\nonumber
\eea
We set $q^2 =0$ and do the integrals over $x$ and $y$ to obtain,
\bea \Gamma_\mu^{(1)I} =
-\frac{2}{(2\pi)^3}\gamma^2\gamma^{\mu'\mu}q_{\mu'}[{\rm
ln}\frac{\Lambda^2}{M^2} - \frac{1}{2}]
\label{9}
\eea
We calculate the term proportional to $M_{\alpha\beta}$ in the same way.
Including this result, we obtain from~(\ref{4}) and~(\ref{9}),
\bea
Z_1^{-1} = 1 - \frac{\gamma^2}{(2\pi)^3}\left( [{\rm ln}\frac{\Lambda^2}{M^2} -
\frac{1}{2}] + \frac{2\gamma^2_{\alpha\beta}}{\gamma^2} [\frac{p'^\alpha
p'^\beta}{M^2} - \frac{(p'^\alpha q^\beta + p'^\beta q^\alpha)}{2M^2} +
\frac{q^\alpha q^\beta}{3M^2}]\right)
\label{10}
\eea
The terms in square brackets represent contributions from new interactions of
the general form,
\bea
\gamma^{\alpha\lambda}\gamma^{\beta}_{\,\,\lambda}\gamma^{\mu\nu}
\phi_a(\partial_\alpha \phi_b)(\partial_\beta\partial_\mu A_\nu).
\nonumber
\eea
This higher-derivative term could become important in the strong field
limit and in non-perturbative regimes.

\subsection{Photon polarization}

Finally, we consider the photon polarization tensor.  We have,
\bea
\Pi_{\mu\nu}(q) = -4iq_\alpha q_\beta \gamma^{\alpha\mu} \gamma^{\beta\nu} \int
\frac{d^4p}{(2\pi)^4}
\frac{1}{(p^2+m_1^2-i\epsilon)((p+q)^2+m_2^2 - i\epsilon)}
\nonumber
\eea
We separate the denominators using the Feynman parameter technique and perform
a Wick rotation as before.  The result is
\bea
\Pi_{\mu\nu}(q) = 4 q_\alpha q_\beta \gamma^{\alpha\mu} \gamma^{\beta\nu}\int
_0^1 dx \int \frac{d^4l}{(2\pi)^4} \frac{1}{(l^2 + M^2)^2}
\nonumber
\eea
where
\bea
M^2 = m_1^2(1-x) + m_2^2x + q^2x(1-x)
\nonumber
\eea
Expanding around the mass shell, we isolate the divergent piece by setting
$q^2=0$, which is equivalent to taking the first term in the expansion.
The result is,
\bea
\Pi_{\mu\nu}(q) = 4 q_\alpha q_\beta \gamma^{\alpha\mu} \gamma^{\beta\nu}
\frac{1}{(2\pi)^3}[{\rm ln}\frac{\Lambda^2}{M^2}+1]
\nonumber
\eea
which leads to an induced interaction of the form,
\bea
\frac{1}{4}\gamma^{\lambda\mu}\gamma^{\tau\nu}F_{\lambda\mu}F_{\tau\nu}
\nonumber
\eea

\subsection{Renormalizability Revisited}

In the introduction it was claimed that a dimensional coupling
constant matrix $\gamma_{\mu\nu}$ ensured the renormalizability of the
model. However, the above calculations show that the one loop
divergences require counter terms of the form
$\gamma^2_{\mu\nu}(\partial^\mu \phi)(\partial^\nu \phi)$ and
$\frac{1}{4}\gamma^{\lambda\mu}\gamma^{\tau\nu}F_{\lambda\mu}F_{\tau\nu}$.
These terms may be thought of as multiplicative modifications to the
scalar field kinetic term in the Lagrangian and to the value of
$\mu_0$ and $\epsilon_0$ in the Maxwell part. They arise because the
orientation of the dipole $\gamma_{\mu\nu}$ breaks the rotational
invariance of the theory, which is then reflected in the quantum
corrections.  We are not obliged to add counterterms of the form
$\gamma^{\alpha\lambda}\gamma^{\beta}_{\,\,\lambda}\gamma^{\mu\nu}
\phi_a(\partial_\alpha \phi_b)(\partial_\beta\partial_\mu A_\nu)$ since
these new interactions yield finite results (at least to one-loop), but
the appearence of such terms nevertheless indicates that they
are an integral part of the structure of the relativistic theory and
should therefore be considered too.

The question then arises: is the theory, as given, renormalizable or
not?  We point out that, in a renormalized field theory, it is the
renormalized values of the parameters which are to be identified with
the physical constants in an experiment.  In our case
$\gamma_{\mu\nu}$ is to be indentified with the dipole moment of an
atomic system.  In fact, the one loop divergences simply tell us that
there are additional, relativistically covariant terms that are second
order in derivatives of the fields that we could have added to the
classical action. These terms correspond to relative permittivities
and permeabilities. These are the only such terms which need to be
added, and with the addition of these terms, the model would indeed be
fully renormalizable. However, note that these two terms involve
$\gamma^2$, which we assume is small, so according to the assumptions
on which we base our perturbative expansion they are probably
negligable. Their physical significance is not redundant however: in
the limit of large electromagnetic fields, very high kinetic energies
and strong dipole couplings, these extra terms become significant and
predict new physics to be identified with experiments.  For the present
paper, we take the pragmatic approach however and assume that such
terms will not contribute signicantly. In effect we are renormalizing
the new $\gamma^2$ couplings to zero. This is a significant
improvement over the non-renormalizable choice of coupling, in which
new, higher derivative interactions would appear at all orders in
perturbation theory.  In renormalization group philosophy, one would
say that we are expanding our theory in a region of Lagrangian-space
which is closer to a renormalization group fixed point.

\section{Conclusions}

We have presented a relativistic model for the interaction of a two
state atom with an electromagnetic field and verified that it reduces
to the Jaynes-Cummings model in the appropriate limits. We have also
shown how to compute higher order, quantum corrections and verified
that the model is one-loop renormalizable. By identifying the
renormalized value of $\gamma_{\mu\nu}$ with observed dipole moments,
or the Rabi flopping frequency of known systems, we have a
prescription for gauging the magnitude of corrections which lead to
the onset of new physics.  The decay rates of the atomic levels may be
identified with imaginary contributions to the self-energy
$\Sigma_{aa}$, for which we are able to calculate an explicit
expression, rather than merely a formal expression as in Korenman's
work. In the present paper, we have been mostly concerned with the
self-consistency of our proposed model and have presented only a
zero-temperature expression for the self-energy. In future work we
shall compute the finite temperature self-energy, where the natural
cut off $\Lambda c^2$ is of the order of $kT$ and obtain a more
accurate gauge of the decay rate by looking at retarded (causal)
boundary conditions, rather than the Feynman boundary conditions used
here. It will also be natural to look at non-equilibrium systems, and
extend our ananlysis to non-linear phenomena where some of the
assumptions made in this paper begin to falter.

Most laser systems are well described by non-relativistic physics.
We consider the most important result of our paper to be the
identification of a model which can be straightforwardly solved in
real-space, with arbitrary boundary conditions, as well as in
many-particle theories at finite temperature and non-equilibrium. The
use of relativistic field theory simplifies calculations greatly
compared to direct non-relativistic formulations.  It also addresses
quantum corrections at the level of the Lamb shift, where corrections
are measurably significant in atomic systems\cite{berkeland1,burgess13},
and removes some of the arbitrariness of previous work on lasers by
tying laser physics to a model which can easily be be analysed within
the framework of a renormalization group philosophy. This is significant
because it indicates which results are independent of the specific
details of microscopic theory one chooses to work with.

Our paper opens a doorway to the study of the statistical mechanics of photons
and atoms in cavities and free space, a topic which we intend to
pursue in later work.  Interesting studies include the use of our
model to study the micromaser with proper finite boundary conditions
and partially reflecting surfaces, and in an expanding or contracting
spherical cavity, as a toy model for light generation by bubbles in
sonoluminescence, and porous silcon.

\section*{Acknowledgments}
We are grateful to David Toms and Cliff Burgess for helpful discussions.
This work is supported by NATO collaborative research grant CRG 950018.


\begin{thebibliography}{10}
\bibitem{jcfoot1}
To be fully covariant, we should write
  $\gamma_{\mu\nu}^\sigma\tilde\epsilon^{{a}{b}}\phi_{a}\partial_\sigma\phi_{b%
}$ where $\sigma$ is normal to a space-like hypersurface, but we suppress the
  extra index for notational simplicity.
\end{thebibliography}

\end{document}